\documentclass[10pt]{article}

\usepackage{axodraw}
\usepackage{amssymb}
\begin{document}

\arraycolsep 2pt
\hfuzz 1.2pt

\renewcommand{\topfraction}{1.0}
\renewcommand{\bottomfraction}{1.0}
\renewcommand{\textfraction}{0.0}

\pagestyle{empty} 
\newcommand{\lpar}{\left(}
\newcommand{\rpar}{\right)} 
\newcommand{\parent}[1]{\lpar#1\rpar}

\newcommand{\lrbr}{\left[}
\newcommand{\rrbr}{\right]}
\newcommand{\rbrak}[1]{\lrbr#1\rrbr}

\newcommand{\ra}{\rightarrow}
\newcommand{\be}{\begin{equation}}
\newcommand{\ee}{\end{equation}}
\newcommand{\bea}{\begin{eqnarray}}
\newcommand{\eea}{\end{eqnarray}}
\newcommand{\beanon}{\begin{eqnarray*}}
\newcommand{\eeanon}{\end{eqnarray*}}
\newcommand{\ba}{\begin{array}}
\newcommand{\ea}{\end{array}}
\newcommand{\bi}{\begin{itemize}}
\newcommand{\ei}{\end{itemize}}
\newcommand{\ben}{\begin{enumerate}}
\newcommand{\een}{\end{enumerate}}
\newcommand{\bc}{\begin{center}}
\newcommand{\ec}{\end{center}}
\newcommand{\ul}{\underline}
\newcommand{\ol}{\overline}
\newcommand{\dotp}{\!\cdot\!}
\newcommand{\nl}{\nonumber \\}

\newcommand{\ord}[1]{{\cal O}\lpar#1\rpar}
\newcommand{\thet}[1]{\theta\lpar#1\rpar}
\newcommand{\delt}[1]{\delta\lpar#1\rpar}

\newcommand{\gtap}{\stackrel{\displaystyle >}{\,_{\! \,_{\displaystyle
\sim}}}}  
\newcommand{\ltap}{\stackrel{\displaystyle <}{\,_{\! \,_{\displaystyle
\sim}}}}  

\newcommand{\eqn}[1]{Eq.(\ref{#1})}
\newcommand{\eqns}[2]{Eqs.(\ref{#1}--\ref{#2})}
\newcommand{\fig}[1]{Fig.~\ref{#1}}
\newcommand{\figs}[2]{Figs.~\ref{#1}--\ref{#2}}
\newcommand{\tab}[1]{Table~\ref{#1}}

\newcommand{\umu}{^{\mu}}
\newcommand{\lmu}{_{\mu}}
\newcommand{\ub}{\bar{u}}
\newcommand{\vb}{\bar{v}}
\newcommand{\gp}{(1+\gamma^5)}
\newcommand{\ga}{\gamma}
\newcommand{\ep}{\epsilon}
\newcommand{\sla}[1]{/\!\!\!#1}
\newcommand{\suml}{\sum\limits}
\renewcommand{\to}{\rightarrow}

\newcommand{\unity}{1\!\!1}

\newcommand{\ssA}{{\scriptscriptstyle \gamma}}
\newcommand{\ssW}{{\scriptscriptstyle W}}
\newcommand{\ssZ}{{\scriptscriptstyle Z}}
\newcommand{\ssX}{{\scriptscriptstyle X}}
\newcommand{\ssB}{{\scriptscriptstyle B}}
\newcommand{\ssV}{{\scriptscriptstyle V}}

\renewcommand{\iff}{\;\;\Longleftrightarrow\;\;}

\newcommand{\stw}{s_w}
\newcommand{\ctw}{c_w}
\newcommand{\stws}{s_w^2}
\newcommand{\stwf}{s_w^4}
\newcommand{\ctws}{c_w^2}
\newcommand{\gw}{g_w}
\newcommand{\gws}{g_w^2}

\newcommand{\Sgg}{\Sigma_\ssA}
\newcommand{\Szg}{\Sigma_\ssX}
\newcommand{\Szz}{\Sigma_\ssZ}
\newcommand{\Sww}{\Sigma_\ssW}

\newcommand{\PW}{$W$}
\newcommand{\PZ}{$Z$}
\newcommand{\gsovermu}{\kappa}

\newcommand{\TeV}{\;\mathrm{TeV}}
\newcommand{\GeV}{\;\mathrm{GeV}}
\newcommand{\MeV}{\;\mathrm{MeV}}

\makeatletter
\ifx\undefined\operator@font
  \let\operator@font=\rm
\fi
\def\Re{\mathop{\operator@font Re}\nolimits}
\def\Im{\mathop{\operator@font Im}\nolimits}

\newcommand{\mut}{m_t^2}
\newcommand{\mw}{m_{_W}}
\newcommand{\gmw}{\Gamma_{_W}}
\newcommand{\mz}{m_{_Z}}
\newcommand{\mzs}{m_{_Z}^2}
\newcommand{\mws}{m_{_W}^2}
\newcommand{\gmz}{\Gamma_{_Z}}

\newcommand\pb{\;[\mathrm{pb}]}

\newcommand{\processccten}{$e^-e^+\to \mu^-\bar{\nu}_\mu u\bar{d}$}
\newcommand{\processcceleven}{$e^-e^+\to s\bar{c} u\bar{d}$}
\newcommand{\processcctwenty}{$e^-e^+\to e^-\bar{\nu}_eu\bar{d}$}


\newcommand{\NP}[1]{{\it Nucl.\ Phys.\ }{\bf #1}}
\newcommand{\PL}[1]{{\it Phys.\ Lett.\ }{\bf #1}}
\newcommand{\ZP}[1]{{\it Z.\ Phys.\ }{\bf #1}}
\newcommand{\NC}[1]{{\it Nuovo Cim.\ }{\bf #1}}
\newcommand{\AP}[1]{{\it Ann.\ Phys.\ }{\bf #1}}
\newcommand{\CMP}[1]{{\it Comm.\ Math.\ Phys.\ }{\bf #1}}
\newcommand{\PR}[1]{{\it Phys.\ Rev.\ }{\bf #1}}
\newcommand{\PRL}[1]{{\it Phys.\ Rev.\ Lett.\ }{\bf #1}}
\newcommand{\MPL}[1]{{\it Mod.\ Phys.\ Lett.\ }{\bf #1}}
\newcommand{\IJMP}[1]{{\it Int.\ J.\ Mod.\ Phys.\ }{\bf #1}}
\newcommand{\JETP}[1]{{\it Sov.\ Phys.\ JETP }{\bf #1}}
\newcommand{\TMP}[1]{{\it Teor.\ Mat.\ Fiz.\ }{\bf #1}}
\newcommand{\HPA}[1]{{\it Helv.\ Phys.\ Acta.\ }{\bf #1}}
\newcommand{\JCP}[1]{{\it Jour. Comp. Phys.\ }{\bf #1}}
\newcommand{\CPC}[1]{{\it Comp. Phys. Comm.\ }{\bf #1}}

\hfuzz 3pt


\begin{flushright}

DFTT-52/99\\
CTP-TAMU-46/99\\
October 1999 
\end{flushright}

\vspace{\fill}

\bc

\vspace{\baselineskip}%
{\Large {\bf Non-conserved currents and gauge-restoring schemes
in single W production}}\\[2em]
\vspace{\baselineskip}%
Elena Accomando$^a$, Alessandro Ballestrero$^{b,c}$ and Ezio Maina$^{b,c}$\\
{\it $^a$ Texas A\&M University, College Station, TX , USA}\\
{\it $^b$INFN, Sezione di Torino, Italy}\\
{\it $^c$Dipartimento di Fisica Teorica, Universit\`a di Torino, Italy}\\
\ec
\vspace{5\baselineskip}
\bc
{\bf Abstract}\\
\ec
We generalize the inclusion of the imaginary parts of the
fermionic one-loop corrections for processes with unstable vector bosons
to the case of massive external fermions
and non conservation of weak currents. 
We study the effect of initial and final state fermion masses in
single \PW production in connection with the gauge-invariant treatment of the
finite-width effects of \PW\ and \PZ\ bosons, 
giving numerical comparisons 
of different gauge-invariance-preserving schemes in
the energy range of LEP2 and LC for \processcctwenty\/.
We do not find significant differences between the results obtained in
the imaginary part fermion loop
scheme and in other exactly gauge preserving methods.

\vspace{\fill}
\newpage
\pagestyle{plain}
\setcounter{page}{1}


\section{Introduction}
\label{se:introduction}

In the past few years a number of papers have discussed
the inclusion of weak boson finite-width effects in the theoretical predictions for
$e^+e^-$ processes. A careful treatment is required since these effects
are intimately related to the gauge invariance of the theory and any violation
of Ward identities can lead to large errors.
Even recently a new proposal for handling unstable particle processes
has appeared \cite{berendschapovsky}. 

The most appealing approach used in actual numerical computations 
is in our opinion the Fermion-Loop (FL) scheme \cite{bhf1, baurzepp, bhf2},
which consists in the 
resummation of the fermionic one-loop corrections to the vector-boson 
propagators and the inclusion of all remaining fermionic one-loop corrections,
in particular those to the Yang--Mills vertices.
In Ref.~\cite{bhf1, baurzepp}  only the imaginary parts of the loops were
included since these represent the minimal set of one--loop contributions which
is required for preserving gauge invariance. This scheme will be referred to
as the Imaginary Part Fermion-Loop (IFL) scheme in the following.
In \cite{bhf2} all contributions from
fermionic one-loop corrections have been computed.
Some effects of light fermion masses in the fermionic loops have been
investigated in \cite{Hoogland&vanoldenborgh}.

In this paper we study the effects of external particle fermion masses
which imply the non--conservation of the weak currents which couple to the
fermionic loops. These effects have been as yet neglected for $e^+ e^-\to
4f$ processes: 
in Ref.~\cite{bhf1} and in the numerical part of Ref.~\cite{bhf2}
all fermions have been assumed to be massless,
while in Ref.~\cite{Hoogland&vanoldenborgh} massive matrix elements
together with the FL corrections of Ref.~\cite{bhf2} were used
under the assumption that the currents were conserved.
Since our main focus is on gauge invariance, we restrict our attention to the
imaginary parts of the fermionic loops, generalizing the approach of
Ref.~\cite{bhf1}.
The extension of the full FL scheme to the case of massive external fermions
is at present being studied \cite{GPnew} and it will allow to determine the
scale of $\alpha_{QED}$ for single \PW\ processes.

We compare the different gauge-restoring schemes in  \processcctwenty\/ (CC20) 
which, in addition to the usual diagrams of  \processccten\/ (CC10), requires
all  diagrams obtained exchanging  the incoming $e^+$ with the outgoing
$e^-$. These contributions become dominant for $\theta_e \ra 0$
because of  the $t$-channel 
$\gamma$ propagator. The CC20 four fermion events with
$e$ lost in the pipe are often referred to as single W production,
and are relevant for triple gauge studies and as background to searches.
For  recent reviews see Refs.~\cite{revs}.
Since the $t$-channel $\gamma$ propagator diverges at $\theta_e=0$ in the
$m_e \ra 0$ limit,
fermion masses have to be exactly accounted for.
Moreover, the apparent $t^{-2}$ behaviour is reduced to $t^{-1}$ by 
gauge cancellations. This implies that even a tiny violation of gauge
conservation can have dramatic effects,
as e.g.\ discussed in Ref.~\cite{bhf1,lmunu}, and the use of some 
gauge conserving scheme is unavoidable.

Two different strategies have been used: 
Improved Weiszacker-Williams \cite{iww} implemented in 
{\tt WTO\cite{wto}} and completely massive codes.
In the first case one separates the 4 $t$-channel photon diagrams, evaluates
them analytically in the equivalent photon approximation taking into account
the complete dependence on all masses, and then adds
the rest of diagrams and the interference between the two sets
in the massless approximation.
In the fully massive MC numerical approach {\tt COMPHEP}\cite{com},
 {\tt GRC4F}\cite{grc}, {\tt KORALW}\cite{kor}, {\tt WPHACT}\cite{wph} and 
recently also the two new codes
{\tt NEXTCALIBUR}\cite{nextc}
and {\tt SWAP\cite{swap}} have  compared their results and found good 
agreement \cite{revs}\footnote{More details can be found in the 
homepage of the  LEP2 MC Workshop 
{\tt http://www.ph.unito.it/\~{}giampier/lep2.html}}
among themselves and with {\tt WTO}.

In the following we first discuss the issue of U(1) gauge invariance
in single W production with non conserved weak currents.
We then give the expression
of all required contributions to the vertex corrections in the IFL scheme.
Finally we present comparisons between the IFL and
other gauge-preserving schemes which have been employed in the literature
and study the relevance of neglecting current non conservation in
the energy range of LEP2 and LC.


\begin{figure}[tb]
\begin{center}
\begin{picture}(90,100)(0,0)
\ArrowLine(10,80)(35,80)
\ArrowLine(35,80)(80,80)
\ArrowLine(80,20)(35,20)
\ArrowLine(35,20)(10,20)
\ArrowLine(80,35)(65,50)
\ArrowLine(65,50)(80,65)
\Photon(35,80)(35,50){2}{4}
\Photon(35,50)(35,20){2}{4}
\Photon(35,50)(65,50){2}{4}
\Vertex(35,80){1.2}
\Vertex(35,20){1.2}
\Vertex(35,50){1.2}
\Vertex(65,50){1.2}
\put(08,90){\makebox(0,0)[r]{$e^-$}}
\put(08,10){\makebox(0,0)[r]{$e^+$}}
\put(82,90){\makebox(0,0)[l]{$e^-$}}
\put(82,10){\makebox(0,0)[l]{$\bar{\nu}_e$}}
\put(82,65){\makebox(0,0)[l]{$u$}}
\put(82,35){\makebox(0,0)[l]{$\bar{d}$}}
\put(30,65){\makebox(0,0)[r]{$\gamma$}}
\put(30,35){\makebox(0,0)[r]{$W$}}
\put(55,55){\makebox(0,0)[b]{$W^+$}}
\end{picture}
\qquad
\begin{picture}(90,100)(0,0)
\ArrowLine(10,80)(35,80)
\ArrowLine(35,80)(80,80)
\ArrowLine(80,20)(55,20)
\ArrowLine(55,20)(35,20)
\ArrowLine(35,20)(10,20)
\ArrowLine(85,35)(70,45)
\ArrowLine(70,45)(85,60)
\Photon(35,80)(35,20){2}{6}
\Photon(55,20)(70,45){2}{3}
\Vertex(35,80){1.2}
\Vertex(35,20){1.2}
\Vertex(55,20){1.2}
\Vertex(70,45){1.2}
\put(08,90){\makebox(0,0)[r]{$e^-$}}
\put(08,10){\makebox(0,0)[r]{$e^+$}}
\put(82,90){\makebox(0,0)[l]{$e^-$}}
\put(82,10){\makebox(0,0)[l]{$\bar{\nu}_e$}}
\put(87,60){\makebox(0,0)[l]{$u$}}
\put(87,35){\makebox(0,0)[l]{$\bar{d}$}}
\put(30,50){\makebox(0,0)[r]{$\gamma$}}
\put(65,32){\makebox(0,0)[br]{$W^+$}}
\end{picture}
\\
\begin{picture}(90,100)(0,0)
\ArrowLine(10,80)(35,80)
\ArrowLine(35,80)(80,80)
\ArrowLine(80,20)(35,20)
\ArrowLine(35,20)(10,20)
\ArrowLine(80,35)(50,35)
\ArrowLine(50,35)(50,65)
\ArrowLine(50,65)(80,65)
\Photon(35,80)(50,65){2}{3}
\Photon(35, 20)(50,35){2}{3}
\Vertex(35,80){1.2}
\Vertex(35,20){1.2}
\Vertex(50,35){1.2}
\Vertex(50,65){1.2}
\put(08,90){\makebox(0,0)[r]{$e^-$}}
\put(08,10){\makebox(0,0)[r]{$e^+$}}
\put(82,90){\makebox(0,0)[l]{$e^-$}}
\put(82,10){\makebox(0,0)[l]{$\bar{\nu}_e$}}
\put(82,65){\makebox(0,0)[l]{$u$}}
\put(82,35){\makebox(0,0)[l]{$\bar{d}$}}
\put(38,65){\makebox(0,0)[r]{$\gamma$}}
\put(42,35){\makebox(0,0)[r]{$W$}}
\end{picture}
\qquad
\begin{picture}(90,100)(0,0)
\ArrowLine(10,80)(35,80)
\ArrowLine(35,80)(80,80)
\ArrowLine(80,20)(35,20)
\ArrowLine(35,20)(10,20)
\ArrowLine(50,35)(80,35)
\ArrowLine(50,65)(50,35)
\ArrowLine(80,65)(50,65)
\Photon(35,80)(50,65){2}{3}
\Photon(35, 20)(50,35){2}{3}
\Vertex(35,80){1.2}
\Vertex(35,20){1.2}
\Vertex(50,35){1.2}
\Vertex(50,65){1.2}
\put(08,90){\makebox(0,0)[r]{$e^-$}}
\put(08,10){\makebox(0,0)[r]{$e^+$}}
\put(82,90){\makebox(0,0)[l]{$e^-$}}
\put(82,10){\makebox(0,0)[l]{$\bar{\nu}_e$}}
\put(82,35){\makebox(0,0)[l]{$u$}}
\put(82,65){\makebox(0,0)[l]{$\bar{d}$}}
\put(38,65){\makebox(0,0)[r]{$\gamma$}}
\put(42,35){\makebox(0,0)[r]{$W$}}
\end{picture}
\end{center}
\vskip -.5cm
\caption[]{The four diagrams of the process $e^-(p_1)e^+(k_1)
\to e^-(p_2)\bar{\nu}_e(k_2) u(p_u)
\bar{d}(p_d)$ which are considered in this paper.}
\label{diagrams_eeenuud}
\end{figure}
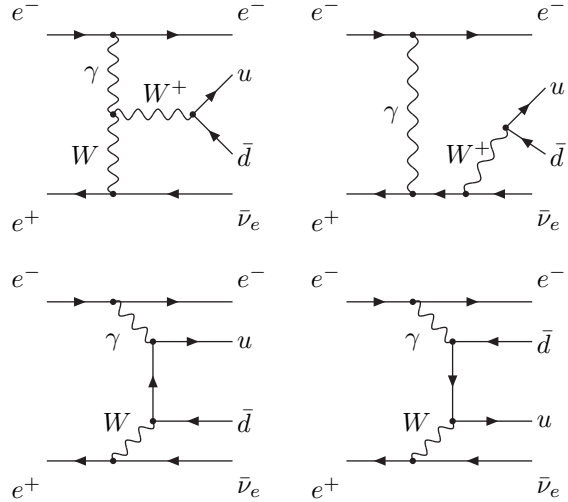

\section{Gauge invariance}
\label{se:gaugeinv}
We choose to work in the unitary gauge. In this case,
the relevant set of Feynman diagrams which become dominant
for $\theta_e \ra 0$ 
coincides  with those discussed in Ref.~\cite{bhf1}.
They are shown in Fig.~\ref{diagrams_eeenuud}.
For ease of comparison we  follow closely the notation of \cite{bhf1}. 
The corresponding  matrix element ${\cal M}$ is given by
\be
{\cal M}
  =
  {\cal M}^{\mu}\;
  J_\mu\;\;,\;\;\;
J^{\mu} = {Q_e \over q^2} \bar{u}(p_2)\gamma^{\mu} u(p_1)\;\;,\;\;\;
  {\cal M}^{\mu}
  =
  \sum_{i=1}^4{\cal M}_i^{\mu}
\ee
where 
\begin{eqnarray}
{\cal M}_1^{\mu}
  & = &
  Q_{_W}\;
  P_{_W}(p_+^2)\;P_{_W}(p_-^2)\;
  V^{\alpha\beta\mu}(p_+,-p_-,-q) 
  D_\alpha^\rho(p_+) D_\beta^\sigma(p_-) {\cal M}^0_{\sigma\rho}\;\;,\nonumber\\
{\cal M}_2^{\mu}
  & = &
  \hphantom{-}
  4iQ_e g_w^2
  \;P_{_W}(p_+^2)\;
  \bar{v}(k_1)\gamma^\mu{\sla{k}_1+\sla{q}-m_e\over(k_1+q)^2-m_e^2}
     \gamma^\alpha P_L v(k_2)\;\;
  \bar{u}(p_u)\gamma_\rho P_L v(p_d)D_\alpha^\rho(p_+)\;\;,\nonumber\\
{\cal M}_3^{\mu}
  & = &
  -4iQ_u g_w^2
  \;P_{_W}(p_-^2)\;
  \bar{u}(p_u)\gamma^\mu{\sla{p}_u-\sla{q}+m_u\over(p_u-q)^2-m_u^2}
    \gamma^\beta  P_L v(p_d)\;\;
     \bar{v}(k_1)\gamma_\sigma  P_L v(k_2)D_\beta^\sigma(p_-)\;\;,\nonumber\\
{\cal M}_4^{\mu}
  & = &
  -4iQ_d g_w^2
  \;P_{_W}(p_-^2)\;
  \bar{u}(p_u)\gamma^\beta P_L {\sla{q}-\sla{p}_d+m_d\over(p_d-q)^2-m_d^2}
    \gamma^\mu v(p_d)\;\;
  \bar{v}(k_1)\gamma_\sigma  P_L v(k_2)D_\beta^\sigma(p_-)\;\;,\nonumber\\
{\cal M}^0_{\sigma\rho}
  & \equiv &
  4i g_w^2\;
  \bar{v}(k_1)\gamma_\sigma  P_L v(k_2)\;\;
  \bar{u}(p_u)\gamma_\rho  P_L v(p_d)\;\;.
\label{fourdiagrams}
\end{eqnarray}
where
$P_L \equiv {1\over2}(1-\gamma^5)$ and
\be
 p_+ = p_u+p_d\;\;,\;\;p_- = k_1-k_2\;\;,\;\;q = p_1-p_2\;\;,
\ee
\be
\label{invprop}
\left[P_{_W}(s)\right]^{-1} = s-M_{_W}^2+i\gamma_{_W}(s)\;\;,
\ee
\be
D_\alpha^\beta(p) = g_\alpha^\beta - p_\alpha
 p^\beta/K(p^2)\;\;.
\ee
$M_{_W}$ is the $W$ mass and
$\gamma_{_W}$ denotes the imaginary part of the inverse $W$ propagator.
 At tree level,
$K(p^2) = M^2_{_W}$ but  the resummation of the imaginary
parts of higher order graphs  modifies the lowest
order expression of $K$ in addition to generate a finite width.
The charged weak coupling constant $g_w$ is given by
$g_w^2 = M_{_W}^2G_F/\sqrt{2}$, while
$Q_i$ is the electric charge of particle $i$, and
\be
V^{\mu_1\mu_2\mu_3}(p_1,p_2,p_3) = (p_1-p_2)^{\mu_3}g^{\mu_1\mu_2} +
(p_2-p_3)^{\mu_1}g^{\mu_2\mu_3} + (p_3-p_1)^{\mu_2}g^{\mu_3\mu_1}\;\;.
\ee
The conservation of electromagnetic current requires
\be
q^\mu {\cal M}_\mu = 0\;\;.
\label{currcons}
\ee
Any small violation of this relation will be amplified 
by a huge factor and will lead to totally wrong predictions
for almost collinear electrons \cite{bhf1,lmunu}.
Multiplying $q^\mu$ into the four diagrams of \eqn{fourdiagrams},
we obtain
\begin{eqnarray}
\label{qdot4diag}
W & \equiv & q^\mu{\cal M}_\mu \nonumber\\
  & = & {\cal M}_0\left\{
  (p_+^2-p_-^2)Q_{_W}\;P_{_W}(p_+^2)\;P_{_W}(p_-^2)\right.\nonumber\\
& & \hphantom{{\cal M}_0}       
   \left.\mbox{}
   +Q_e\;P_{_W}(p_+^2) - \left( Q_d-Q_u\right)\;
     P_{_W}(p_-^2) \right\}\;\;\nonumber\\
& &  -{\cal M}_{++}\left\{
        Q_{_W}\;P_{_W}(p_+^2)\;P_{_W}(p_-^2)\;
        \left( 1 - p_-^2/K(p_+^2)\right)
        \right.\nonumber\\
& & \hphantom{{\cal M}_{++}\;}
   \left.\mbox{}
         +Q_e\;P_{_W}(p_+^2)\;/K(p_+^2)
        \right\}
        \\
& &  +{\cal M}_{--}\left\{
        Q_{_W}\;P_{_W}(p_+^2)\;P_{_W}(p_-^2)\;
        \left( 1 - p_+^2/K(p_-^2)\right)
        \right.\nonumber\\
& & \hphantom{{\cal M}_{--}\;}
   \left.\mbox{}
        + \left( Q_d-Q_u\right)\;P_{_W}(p_-^2)\;/K(p_-^2)
        \right\}
        \nonumber\\
& &  +{\cal M}_{-+}\left\{
        Q_{_W}\;P_{_W}(p_+^2)\;P_{_W}(p_-^2)\; p_-\cdot p_+\;
        \left(
        K(p_-^2)^{-1} - K(p_+^2)^{-1} \right)
        \right\}\;.
        \nonumber
\end{eqnarray}

\newpage

where
\be
{\cal M}_0
  \equiv 
  {\cal M}^0_{\alpha\beta}\,g^{\alpha\beta}\;,\; \;\;\;\;\;\;\;
{\cal M}_{++} 
   \equiv 
  {\cal M}^0_{\alpha\beta}\,p_+^\alpha p_+^\beta\;,
\ee 
\be
{\cal M}_{--} 
  \equiv 
  {\cal M}^0_{\alpha\beta}\,p_-^\alpha p_-^\beta\;,\; \;\;\;
{\cal M}_{-+} 
  \equiv
  {\cal M}^0_{\alpha\beta}\,p_-^\alpha p_+^\beta\;\;.
\ee

Using $Q_{_W} = Q_e = Q_d-Q_u = - 1$ and \eqn{invprop} we have
\begin{eqnarray}
\label{gaugecancellation1}
W & = &\; i\;{\cal M}_0\;P_{_W}(p_+^2)\;P_{_W}(p_-^2)\;
        \left(\gamma_{_W}(p_+^2)-\gamma_{_W}(p_-^2)\right) \nonumber\\
& &  +\;{\cal M}_{++}\left\{
        P_{_W}(p_+^2)\;P_{_W}(p_-^2)\;
        \left( 1 - \parent{M_{_W}^2-i\gamma_{_W}(p_-^2)}/K(p_+^2)\right)
        \right\}  \\
& &  -\;{\cal M}_{--}\left\{
        P_{_W}(p_-^2)\;P_{_W}(p_+^2)\;
        \left( 1 - \parent{M_{_W}^2-i\gamma_{_W}(p_+^2)}/K(p_-^2)\right)
        \right\} \nonumber\\
& &  -\;{\cal M}_{-+}\left\{
        P_{_W}(p_+^2)\;P_{_W}(p_-^2)\; p_-\cdot p_+\;
        \left( K(p_-^2)^{-1} - K(p_+^2)^{-1} \right)
        \right\}\;.
        \nonumber
\end{eqnarray}
Current conservation is therefore violated unless
\begin{eqnarray}
\gamma_{_W}(p_+^2) &=& \gamma_{_W}(p_-^2) \equiv \ol{\gamma}_{_W}\\
K(p_+^2) &=& K(p_-^2) = M_{_W}^2-i\ol{\gamma}_{_W}
\label{fixwidthcondition} 
\end{eqnarray}
It should be mentioned that all effects due to the non conservation
of the currents
which couple to the $W$ and $Z$ bosons are contained in the last three terms of 
\eqn{qdot4diag} and  \eqn{gaugecancellation1} which would be zero if the
currents were conserved.

The most naive treatment of a
Breit-Wigner resonance uses a {\em fixed width\/} approximation, with
\begin{equation}
\ol{\gamma}_{_W} = M_{_W}\Gamma_{_W}\;\;.
\end{equation}

\eqns{gaugecancellation1}{fixwidthcondition}
show that in this case there is no violation of
electromagnetic current conservation. 
In the unitary gauge this corresponds to adding the same imaginary part,
$-iM_{_W}\Gamma_{_W}$, to $M^2_{_W}$ both in the denominator and in the 
$p^\mu p^\nu$ term of the \PW propagator (see {\it e.g.} \cite{baurzepp}
and references therein). We have verified numerically
that neglecting to modify the latter leads to large errors already at
$800 \GeV$. 
A similar approach, in which 
all weak boson masses squared $M^2_{_B}\;, B = W,Z$ are changed to
$M^2_{_B}-i\gamma_{_B}$ everywhere, 
including in the definition of the weak mixing angle,
has in fact been suggested \cite{DDRW}
as a mean of preserving both U(1) and SU(2) Ward identities in
the Standard Model.

The fixed-width approximation cannot however
be justified from field theory. Indeed, 
propagators with space-like momenta are real and cannot
acquire a finite width in contradiction to the fixed-width scheme.

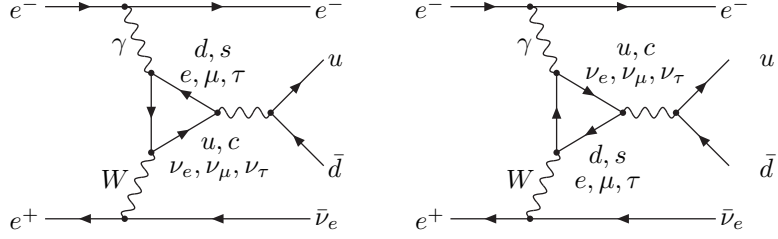
\begin{figure}[tb]
\begin{center}
\begin{picture}(130,120)(0,0)
\ArrowLine(10,100)(40,100)
\ArrowLine(40,100)(110,100)
\ArrowLine(110,20)(40,20)
\ArrowLine(40,20)(10,20)
\ArrowLine(115,40)(95,60)
\ArrowLine(95,60)(115,80)
\Photon(40,100)(50,75){2}{4}
\Photon(40, 20)(50,45){2}{4}
\ArrowLine(50,75)(50,45)
\ArrowLine(50,45)(75,60)
\ArrowLine(75,60)(50,75)
\Photon(75,60)(95,60){2}{3}
\Vertex(40,100){1.2}
\Vertex(40,20){1.2}
\Vertex(50,45){1.2}
\Vertex(50,75){1.2}
\Vertex(95,60){1.2}
\Vertex(75,60){1.2}
\put(08,100){\makebox(0,0)[r]{$e^-$}}
\put(08,20){\makebox(0,0)[r]{$e^+$}}
\put(112,100){\makebox(0,0)[l]{$e^-$}}
\put(112,20){\makebox(0,0)[l]{$\bar{\nu}_e$}}
\put(117,80){\makebox(0,0)[l]{$u$}}
\put(117,40){\makebox(0,0)[l]{$\bar{d}$}}
\put(41,85){\makebox(0,0)[r]{$\gamma$}}
\put(42,35){\makebox(0,0)[r]{$W$}}
\put(61,70){\makebox(0,0)[bl]{\shortstack{$d,s$\\$e,\mu,\tau$}}}
\put(57,51){\makebox(0,0)[tl]{\shortstack{$u,c$\\$\nu_e,\nu_\mu,\nu_\tau$}}}
\end{picture}
\qquad
\begin{picture}(130,120)(0,0)
\ArrowLine(10,100)(40,100)
\ArrowLine(40,100)(110,100)
\ArrowLine(110,20)(40,20)
\ArrowLine(40,20)(10,20)
\ArrowLine(115,40)(95,60)
\ArrowLine(95,60)(115,80)
\Photon(40,100)(50,75){2}{4}
\Photon(40, 20)(50,45){2}{4}
\ArrowLine(50,45)(50,75)
\ArrowLine(75,60)(50,45)
\ArrowLine(50,75)(75,60)
\Photon(75,60)(95,60){2}{3}
\Vertex(40,100){1.2}
\Vertex(40, 20){1.2}
\Vertex(50, 45){1.2}
\Vertex(50, 75){1.2}
\Vertex(95,60){1.2}
\Vertex( 75,60){1.2}
\put(08,100){\makebox(0,0)[r]{$e^-$}}
\put(08,20){\makebox(0,0)[r]{$e^+$}}
\put(112,100){\makebox(0,0)[l]{$e^-$}}
\put(112,20){\makebox(0,0)[l]{$\bar{\nu}_e$}}
\put(127,80){\makebox(0,0)[l]{$u$}}
\put(127,40){\makebox(0,0)[l]{$\bar{d}$}}
\put(41,85){\makebox(0,0)[r]{$\gamma$}}
\put(42,35){\makebox(0,0)[r]{$W$}}
\put(61,70){\makebox(0,0)[bl]{\shortstack{$u,c$\\$\nu_e,\nu_\mu,\nu_\tau$}}}
\put(57,48){\makebox(0,0)[tl]{\shortstack{$d,s$\\$e,\mu,\tau$}}}
\end{picture}
\end{center}
\vskip -1cm
\caption[]{The extra fermionic diagrams needed to cancel the gauge-breaking
terms.}
\label{extra_diagrams_eeenuud}
\end{figure}

As discussed in Ref.~\cite{bhf1},
the simplest way to restore gauge-invariance
in a theoretically satisfying fashion is the addition of the imaginary parts
of one-loop fermionic vertex corrections,
shown in Fig.~\ref{extra_diagrams_eeenuud}, which cancel the imaginary
part in the Ward identities.
The cancellation is exact as long as all fermion loops, both in the vertices and
in the propagators, are computed in the same approximation. 
In particular we can 
consistently neglect fermion masses in the loops,
if we use 
for the $W$ width
the tree--level expression for the decay of an on-shell $W$ to massless fermions
\begin{equation}
\label{widthw}
\Gamma_{_W} = \sum_{\mbox{\tiny doublets}}\;N_f\;
  {G_FM_{_W}^3\over6\pi\sqrt{2}}\;\;,
\end{equation}
involving a sum over all fermion doublets with $N_f$ (1 or 3)
colours.

The vertex corrections are given by


\begin{eqnarray}
\label{eq:M5}
{\cal M}_5^{\mu}   &=&  {i\over16\pi}{\cal M}^0_{\rho\sigma}\;P_{_W}(p_+^2)\;
    P_{_W}(p_-^2)\;g_w^2 \times \\
  & & \hphantom{--------} \times \sum_{\mbox{\tiny doublets}}N_f\left( Q_d-Q_u \right)\;
       D_\alpha^\rho(p_+) D_\beta^\sigma(p_-)\; Z^{\alpha\beta\mu}\;\;, \nonumber
\end{eqnarray}
where 
\begin{equation}
\label{eq:Z1}
Z^{\alpha\beta\mu} = {1\over2\pi} \int d\Omega
     \;\mbox{Tr}\left[ \sla{r}_1\gamma^\mu{\sla{r}_1-\sla{q}\over(r_1-q)^2}
                         \gamma^\beta\sla{r}_2\gamma^\alpha
                  \right]
\end{equation}
 is the imaginary part of the triangle insertions. The momenta $r_1$ and
$r_2$ are the momenta of the cut fermion lines with $p_+=r_1+r_2$.
The expression $Z^{\alpha\beta\mu}$ satisfies the three Ward
identities:
\begin{eqnarray}
\label{wiz}
Z^{\alpha\beta\mu} q_\mu
  & = & -{8\over3}\left( p_+^\alpha p_+^\beta - p_+^2 g^{\alpha\beta}
                                         \right) \;\;,\nonumber\\
Z^{\alpha\beta\mu} p^+_\alpha
  & = & 0 \;\;,\\
Z^{\alpha\beta\mu} p^-_\beta
  & = & +{8\over3}\left( p_+^\mu p_+^\alpha - p_+^2 g^{\mu\alpha}
                                        \right) \;\;.\nonumber
\end{eqnarray}
Because of the anomaly cancellation we have no explicit contributions from the
part containing $\gamma^5$.
Attaching the photon momentum $q_\mu$ to the sum of the
diagrams ${\cal M}_5^{\mu}$ gives
\begin{eqnarray}
W_{\mathrm{add}}  \equiv  q_\mu{\cal M}_5^{\mu} 
  & = &
   - \; i \;{\cal M}_0\;P_{_W}(p_+^2)\;P_{_W}(p_-^2)\;
  \Gamma_{_W} {p_+^2\over M_{_W}}\nonumber\\
 & & +\; i\;{\cal M}_{++}\;P_{_W}(p_+^2)\;P_{_W}(p_-^2)\;
         {\Gamma_{_W} \over M_{_W}} \\
 & & +\; i\;{\cal M}_{--}\;P_{_W}(p_+^2)\;P_{_W}(p_-^2)\;       
        {\Gamma_{_W} \over M_{_W}}{p_+^2\over K(p_-^2)}
       \nonumber\\
 & & -\; i\;{\cal M}_{-+}\;P_{_W}(p_+^2)\;P_{_W}(p_-^2)\;       
        {\Gamma_{_W} \over M_{_W}}
        {{p_+\cdot p_-}\over K(p_-^2)}
       \nonumber
\label{qdotM5}       
\end{eqnarray}
where we used the Ward identity of \eqn{wiz} and the definition of the
nominal $W$ width, \eqn{widthw}.
\begin{figure}[tb]
\begin{picture}(200,30)(0,0)
\Photon(110,15)(130,15){2}{4}
\put(122,20){\makebox(0,0)[b]{$W$}}
\BCirc(140,15){10}
\Photon(150,15)(170,15){2}{4}
\put(162,20){\makebox(0,0)[b]{$W$}}
\put(175,15){\makebox(0,0)[l]
       {$= - i \,\Pi^{\mu\nu}_{\ssW}$}}
\end{picture}
\vskip -.3cm
\caption[]{First order contribution to the inverse W propagator.
The fermions in the loop are assumed to be massless.}
\label{invWprop}
\end{figure}
Assuming  $\gamma_{_W}(p_-^2)=0$ 
as required by field theory, the extra diagrams restore U(1) gauge invariance
provided
\begin{eqnarray}
\gamma_{_W}(p_+^2) &=& \Gamma_{_W} {p_+^2\over M_{_W}}\; , \\
K(p_+^2) &=& M^2_{_W} \left( 1 + i {\Gamma_{_W} \over M_{_W}}\right)^{-1}\;,\\
K(p_-^2) &=& M^2_{_W}\;\;.
\label{fixedW}
\end{eqnarray}
This result may be surprising but it is actually the correct field theoretical
resummation, in the unitary gauge, of the imaginary part of the fermionic 
one--loop contributions \cite{baurzepp},
shown in \fig{invWprop}, which is transverse if we consider only 
massless fermions 
\be
\Im \left( \Pi^{\mu\nu}_{\ssW}\right) = 
\left( g^{\mu\nu} - p^\mu p^\nu/p^2 \right)\Pi_{\ssW}
\ee
with
\be
\Pi_{\ssW} = p^2 {\Gamma_{_W} \over M_{_W}}\;\;.
\label{treeWwidth}
\ee

If, suppressing indices for simplicity, we define $\unity \equiv g^{\mu\nu}$,
$D \equiv g^{\mu\nu} - p^\mu p^\nu/M^2$ and 
$T \equiv g^{\mu\nu} - p^\mu p^\nu/p^2$ we have $DT = TD = T$ and $T^2 = T$.
The usual Dyson series for the resummation of the imaginary part $\Pi$
of one--loop corrections reads:
\bea
S & = & {-iD \over p^2- M^2} + {-iD \over p^2- M^2}(T\Pi){-iD \over p^2- M^2} +
\cdots \nl
& = &  {-iD \over p^2- M^2 +i\,\Pi}\left( \unity + {i\,\Pi \over p^2-M^2}\left(\unity-T\right)
        \right) 
\eea
More explicitly
\bea
S^{\mu\nu} 
        & = & {-i \over p^2- M^2 +i\Pi}
        \left\{
         g^{\mu\nu} 
         - {p^\mu p^\nu \over M^2}
               \left(1 + {i\Pi \over p^2}\right)\right\} 
\eea

Hence the introduction of a finite width for $s$-channel
virtual \PW 's which is required even in tree level calculations
has to be associated with a corresponding modification of
the $p^\mu p^\nu$ term.
 
\newpage

\par
\section{Form factors for the vertex corrections}
\label{se:formfactors}
We report in this section the analytic expression of $Z^{\alpha\beta\mu}$
which is needed for actual computations in the FL scheme.
Parametrizing $Z^{\alpha\beta\mu}$ as follows                                       
\bea
\label{ourZ}
Z^{\alpha\beta\mu} &=& \;\; p_+^\alpha p_+^\beta p_+^\mu f_1
                    + q^\alpha p_+^\beta p_+^\mu f_2
                    + p_+^\alpha q^\beta p_+^\mu f_3
                    + p_+^\alpha p_+^\beta q^\mu f_4  \nl
                 & &   + q^\alpha q^\beta p_+^\mu f_5
                    + q^\alpha p_+^\beta q^\mu f_6
                    + p_+^\alpha q^\beta q^\mu f_7
                    + q^\alpha q^\beta q^\mu f_8 \\
                 & &   + g^{\alpha\beta} p_+^\mu f_9
                    + g^{\alpha\beta} q^\mu f_{10}
                    + g^{\beta\mu} p_+^\alpha f_{11}
                    + g^{\beta\mu} q^\alpha f_{12} \nl
                 & &   + g^{\alpha\mu} p_+^\beta f_{13}
                    + g^{\alpha\mu} q^\beta f_{14} \nonumber
\eea
we find
\bea
 f1 &=&                                                                     
    160 \frac{p_+^2 q^2 p_-^2}{\lambda^3} 
    \left\{ 
       f_0 \left[ -6 p_+^2 q^2 p_-^2 -2 ( p_+^2 + p_-^2 )                    
     ({p_- \cdot p_+})^2 + 2 ( p_+^4 +  p_-^4 ){p_- \cdot p_+}  \right]                              
         \rule[-.3 cm]{0cm}{.8cm} \right.   \nl
     & &        +  10 \frac{p_+^2 ({p_- \cdot p_+})^2}{q^2}              
     + 20 p_+^2 p_-^2 + 10 p_+^4 
     - \frac{2}{q^2} \left( 3 p_+^6                    
        + ({p_- \cdot p_+})^2 p_-^2 + p_-^6  \right)                 
        + 2 p_-^4     \nl
   & &    + \frac{\lambda}{10} \left[ f_0 \left( - \frac{q^2 p_-^2}{p_+^2}+               
           \frac{p_-^4}{p_+^2}- \frac{p_+^2 q^2}{p_-^2} - 6 p_+^2                             
         + \frac{p_+^4}{p_-^2} - 10 q^2 - 6 p_-^2 \right)  
      \rule[-.3 cm]{0cm}{.8cm} \right.   \\                        
   & &     
      \left.  \rule[-.3 cm]{1.cm}{0.cm} 
       \frac{116}{3}- 2 \frac{p_-^4}{q^2 p_+^2} 
       + 2 \frac{p_-^2}{p_+^2} + \frac{139}{3} \frac{p_+^2}{q^2}               
          + 14 \frac{p_+^2}{p_-^2} 
       - \frac{20}{3} \frac{p_+^4}{q^2 p_-^2}
       + \frac{67}{3} \frac{p_-^2}{q^2}                                  
      \rule[-.3 cm]{0cm}{.8cm} \right]   \nl
   & &    
    \left.  + \frac{\lambda^2}{10} \left[ 
                 - f_0 \left( \frac{1}{p_+^2} + \frac{1}{p_-^2} \right) 
            + \frac{7}{3 p_+^2 q^2} + \frac{2}{3 p_+^2 p_-^2}
            + \frac{19}{3 q^2 p_-^2} \right]
      \right\}\;,  \nl 
     & & \nl
     & & \nl
f2  &=&
         -160 \frac{p_+^2 q^2 p_-^2}{\lambda^3} 
            \left\{   
      f_0 \left[ p_+^2 q^2 p_-^2 + p_+^2 q^4 - q^2 p_-^4 +2 q^4 p_-^2 - q^6
                 \right]
      \rule[-.3 cm]{0cm}{.8cm} \right.   \nl
     & &   +  2 p_+^2 {q\cdot p_-} +6 q^2 {q\cdot p_-} 
               +4 q^4 -2 {q\cdot p_-} p_-^2   \nonumber     \\  
     & & + \frac{f_0 \lambda}{20} \left( 2 \frac{p_+^2 q^2}{ p_-^2} + 3 p_+^2 +17 q^2 
                -2 \frac{q^4}{p_-^2} -3 p_-^2 \right) \\
     & & + \left.
       \frac{\lambda}{15 p_-^2}( 11 q^2 + 13{q\cdot p_-} - p_-^2 )
        + \frac{\lambda^2}{60 q^2 p_-^2}( 1 + 3 f_0 q^2 )
           \right\}\;, \nl
     & & \nl
     & & \nl
f3 &=&
    160 \frac{p_+^2 q^2 p_-^2}{\lambda^3} 
    \left\{ 
           f_0   \left[ 3 p_+^2 q^2 p_-^2 + (3 p_+^2 - p_-^2 )({p_-\cdot p_+})^2
                 - 2 p_+^4 {p_-\cdot p_+}  \right]
      \rule[-.3 cm]{0cm}{.8cm} \right.   \nl
      & &         - 8 \frac{p_+^2 ({p_-\cdot p_+})^2}{ q^2} 
             - 8 p_+^2 p_-^2 - 8 p_+^4 
            + 4 \frac{p_+^6}{q^2}
            + 4 \frac{({p_-\cdot p_+})^2 p_-^2}{q^2} 
        \nl
       & &    + \frac{\lambda}{10} \left[    
            f_0  \left( \frac{p_+^2 q^2}{p_-^2} + 3 p_+^2 
            - \frac{p_+^4}{p_-^2} + \frac{13}{2} q^2 + 3 p_-^2 \right)
      \rule[-.3 cm]{0cm}{.8cm} \right.   \nl
      & &   \left. \rule[-.3 cm]{1.cm}{0.cm} 
      - 20 - 32 \frac{p_+^2}{q^2} - \frac{40}{3}\frac{ p_+^2}{ p_-^2}
      + 6 \frac{p_+^4}{q^2 p_-^2} - 4 \frac{p_-^2}{q^2} 
               \right]   \\
   & & 
       \left. 
           + \frac{\lambda^2}{20}
            \left[ f_0 \left( \frac{1}{ p_+^2} + \frac{2}{ p_-^2} \right)
                - \frac{1}{ p_+^2 q^2} - \frac{1}{p_+^2 p_-^2}
            - \frac{34}{3 q^2 p_-^2} \right]
      \rule[-.3 cm]{0cm}{.8cm} \right\}\;,   \nl
     & & \nl
     & & \nl
f5 &=&
         160 \frac{p_+^2 q^2 p_-^2}{\lambda^3} 
            \left\{   
             f_0 \left[
        3 p_+^2 q^2 p_-^2 + p_+^2 q^4 - q^2 p_-^4 + 2 q^4 p_-^2 - q^6 \right]  
      \rule[-.3 cm]{0cm}{.8cm} \right.   \nl
     & &  + 6 p_+^2 {q\cdot p_-} + 6 q^2 {q\cdot p_-} + 4 q^4 - 2 {q\cdot p_-}p_-^2 
      \nl    
     & & + \frac{f_0 \lambda}{20}( 2 \frac{p_+^2 q^2}{ p_-^2} + 7 p_+^2 + 21 q^2
         - 2 \frac{q^4}{ p_-^2} + p_-^2 )      \\    
     & & \left .
       + \frac{\lambda}{5}\left( - 3 + \frac{2}{3} \frac{q\cdot p_-}{q^2} 
             + \frac{11}{3}\frac{q^2}{ p_-^2} + 5 \frac{{q\cdot p_-}}{ p_-^2} 
                    \right)              
        + \frac{\lambda^2}{60 q^2 p_-^2}( 1 + 3 f_0 q^2 )
           \right\}\;,   \nl
     & & \nl
     & & \nl
f9 &=&
         16 \frac{p_+^2 q^2 p_-^2}{\lambda^2} 
            \left\{   
           f_0   \left[ q^2 {p_-\cdot p_+} + \frac{\lambda}{2}\right]
             + 2 {q\cdot p_+}
               + \frac{\lambda}{6 q^2 p_-^2} ( 4{p_+\cdot q} - 3 q^2 )
            \right\}  \\
     & & \nl
     & & \nl
f11 &=&
         16 \frac{p_+^2 q^2 p_-^2}{\lambda^2} \left\{ 
           f_0 \left[ q^2 {p_-\cdot p_+} + \lambda \left( - \frac{1}{4} 
             +\frac{{p_-\cdot p_+}}{2 p_+^2}  \right) \right] \right.  \\
     & &  \left.     + 2 {p_+\cdot q} 
            + \lambda  {p_+\cdot q} (\frac{1}{2 p_+^2 q^2} + 
       \frac{1}{ 2 p_+^2 p_-^2} -\frac{2}{3 q^2p_-^2 } ) \right\}\;,
    \nl
     & & \nl
     & & \nl
f12 &=&
         16 \frac{p_+^2 q^2 p_-^2}{\lambda^2} \left\{ 
           f_0 \left[- p_+^2 {q\cdot p_-} + \frac{\lambda}{2}\right] 
          - 2 p_+^2 
            + \frac{\lambda}{6 q^2 p_-^2} ( 6{q\cdot p_-} - p_+^2)  \right\}\;,
          \\
     & & \nl
     & & \nl
f13 &=&
         16 \frac{p_+^2 q^2 p_-^2}{\lambda^2} \left\{ 
         \phantom{\left( - \frac{3}{4}\right.} \right. \nl
    & &     \left.
         f_0 \left[ - 7 q^2{p_+\cdot q}  
         + 4 ({p_+\cdot q})^2 - 3 q^2 p_-^2 + 3 q^4  
           + \lambda   \left( - \frac{3}{4} + \frac{{p_+\cdot q}}{2 p_-^2}
              -\frac{q^2}{2 p_-^2} \right)  \right]  \right. \\
 & &  \left.   + \frac{8}{3}( p_-^2 -  q^2 ) - \frac{22}{3}{p_-\cdot p_+} 
           + \frac{14}{3}\frac{({p_-\cdot p_+})^2}{ p_-^2}
               + \frac{\lambda}{6 q^2 p_-^2}  {q\cdot p_-}  
       \right\}\;,          \nl
     & & \nl
     & & \nl
f14 &=&
         16 \frac{p_+^2 q^2 p_-^2}{\lambda^2} \left\{ 
            f_0 \left[ 7 {p_+\cdot q} q^2 -{p_+\cdot q} p_-^2 - 4 ({p_+\cdot q})^2 
            + 3 q^2  p_-^2 - 3 q^4 \rule[-.3 cm]{0cm}{.8cm} \right. 
      \rule[-.3 cm]{0cm}{.8cm} \right.   \nl
     & &  \left.    +\lambda   \left( \frac{1}{4} - \frac {p_+\cdot q} {2 p_-^2} 
                    + \frac{q^2}{2 p_-^2} \right)  \right] 
- \frac{8}{3}( p_-^2 -  q^2 )  + \frac{16}{3} {p_-\cdot p_+} 
            - \frac{14}{3}\frac{({p_- \cdot p_+})^2}{ p_-^2}    \\
    & &  \left.  
               + \frac{\lambda}{6 q^2 p_-^2} (  4 p_-^2 - 5{p_-\cdot p_+} )
\right\}\;\;. \nonumber
\label{formfacts}
\eea
with
\be  
f_0 = -{2\over \sqrt{\lambda}}\ln\left({2(p_-\!\!\cdot q)+\sqrt{\lambda}
          \over 2(p_-\!\!\cdot q)-\sqrt{\lambda}}\right)\;, \hphantom{---}
\lambda  \equiv
         4(p_-\!\!\cdot q)^2-4p_-^2q^2\;\;.
\ee 

The form factors $f_4$, $f_6$, $f_7$, $f_8$, $f_{10}$ do not contribute
in CC20 processes because the electron current is conserved.
If we assume that all currents which couple to the fermion loops are conserved
we have
\be
Z^{\alpha\beta\mu} = q^\alpha q^\beta p_+^\mu c_0
                      + g^{\beta\mu} q^\alpha c_1
                      + g^{\alpha\mu} q^\beta c_2
                      + g^{\alpha\beta} p_+^\mu c_3 
\ee
\be 
               c_0 = f_2 + f_5 \hphantom{---}
                c_1 = f_{12} \hphantom{---}
                c_2 = f_{13} +  f_{14} \hphantom{---}
                c_3 = f_9 
\ee
where the $c_i$ agree with the results of Ref.\cite{bhf1}.

\section{Applications to the process \processcctwenty: numerical effects in a physically
relevant case study}
\label{se:numerics}

In the following we present numerical results obtained with the IFL
scheme and make comparisons with those obtained with other gauge-preserving 
approaches.
 The following schemes are considered in our analysis:
\begin{description}

\item[Imaginary-part FL scheme(IFL):] 
    The imaginary part of the fermion-loop corrections 
    \eqns{ourZ}{formfacts}
     are used. The fermion masses  are neglected in the loops but not in the
    rest of the diagrams. 

\item[Fixed width(FW):]  All \PW-boson propagators are given by 
\be 
{g^{\mu\nu}-{p^\mu p^\nu \over M^2_{_W} -i\Gamma_{_W}M_{_W}}
\over p^2-M^2_{_W}+i\Gamma_{_W}M_{_W}}\;\;.
\ee
   This gives an unphysical width for $p^2<0$, but retains  U(1) gauge 
   invariance.

\item[Complex Mass(CM):]
All weak boson masses squared $M^2_{_B}\;, B = W,Z$ are changed to
$M^2_{_B}-i\gamma_{_B}$, including when they appear 
in the definition of the weak mixing angle. This scheme has the advantage of
preserving both U(1) and SU(2) Ward identities \cite{DDRW}.
    
\item[Overall scheme(OA):]
    The diagrams for \processcctwenty\ can be split into two sets which are
    separately gauge invariant under U(1). 
    In the present implementation of OA \cite{overall}, $t$-channel diagrams
    are computed without any width and are then multiplied
    by $(q^2-M^2)/(q^2-M^2+iM\Gamma)$ where $q$, $M$ and $\Gamma$ are the
    momentum, the mass and
    the width of the possibly-resonant \PW-boson.
    This scheme  retains U(1) gauge invariance  at the
    expenses of mistreating  non resonant terms.
    
\end{description}

In order to assess the relevance of current non-conservation in
process \processcctwenty\ we have also implemented
the imaginary part of the fermion-loop corrections with the
assumption that all currents which couple to the fermion-loop are conserved.
In this case  \eqns{ourZ}{formfacts} reduce to those
computed in ~\cite{bhf1}.
Notice that the masses of external fermions are nonetheless taken into account
in the calculation of the matrix elements. This scheme violates  
U(1) gauge invariance
by terms which are proportional to the fermion masses squared, as already
noted in Ref.~\cite{Hoogland&vanoldenborgh}. However they are enhanced 
at high energy by large factors and can be numerically quite relevant.
This scheme will be referred to as the 
imaginary-part FL scheme with conserved currents {\bf (IFLCC)}
in the following.

\begin{table}[tb]\centering
\begin{tabular}{|c|c|c|c|} 
\hline
            & 190 GeV         & 800 GeV         & 1500 GeV           \\
\hline
IFL       &   0.11815 (13)   &   1.6978 (15)   &    3.0414 (35)     \\
\hline
FW        &   0.11798 (11)   &   1.6948 (12)   &    3.0453 (41)     \\
\hline
CM        &   0.11791 (12)  &  1.6953 (16)  &  3.0529  (60)   \\
\hline
OA        &   0.11760 (10)   &   1.6953 (13)   &    3.0401 (23)     \\   
\hline
\hline
IFLCC     &   0.11813 (12)   &   1.7987 (16)   &    5.0706 (44)     \\
\hline
\end{tabular}
\caption{Cross sections for the process $e^+ e^-~\ra\; e^- \bar
\nu_e u\bar d$ for various gauge restoring schemes}
\label{tab1}
\end{table}
In the comparisons among the different codes mentioned in the introduction,
{\tt COMPHEP} and {\tt WPHACT} used the OA scheme, {\tt KORALW} and {\tt GRC4F}
 the $L^{\mu\nu}$ transform method of Ref.~\cite{lmunu}, 
{\tt NEXTCALIBUR} used the CM and  {\tt SWAP} the FW  scheme.
Here all schemes described above  have been implemented in the new version 
of {\tt WPHACT}
in which all massive matrix elements have been added to the old massless ones.
In particular the IFL contributions in \eqns{ourZ}{formfacts} have been 
introduced.
In this way, the same matrix elements, phase spaces and integration routines
are used in all instances.
\begin{table}[bt]\centering
\begin{tabular}{|c|c|c|c|} 
\hline
      & $\cos (\theta_e) >$ .997 &  
                        $\theta_e <$ 0.1 degree & 
                                            M$(u \bar d) >$ 40  GeV  \\
\hline
IFL        &   1.6978 (15)    &   1.1550 (15)    &   1.6502 (15)  \\
\hline
FW         &   1.6948 (12)    &   1.1538 (21)    &   1.6480 (13)  \\
\hline
CM         &   1.6953 (16)    &   1.1533 (14)    &   1.6520 (10)   \\
\hline
OA         &   1.6953 (13)   &   1.1537 (12)    &   1.6523 (12)  \\     
\hline 
\hline
IFLCC      &   1.7987 (16)   &   1.2600 (22)   &    1.7424 (21)     \\
\hline
\end{tabular}
\caption{Cross sections for the process $e^+ e^-~\ra\; e^- \bar
\nu_e u\bar d $ at E=800 GeV for various gauge restoring schemes
and different cuts.}
\label{tab2}
\end{table}

If not stated otherwise we apply the following cuts:
\be
M(u \bar d) > 5 \GeV\;, E_u > 3 \GeV, E_{\bar d} > 3 \GeV, \cos (\theta_e ) > .997
\ee
We have produced numerical results for \processcctwenty\ in the
small space-like $q_\ga^2$ (collinear electron) region 
where we expect gauge-invariance issues to be essential. 
We have not included in our computations Initial State Radiation (ISR),
in order to avoid any additional uncertainty in these comparisons among
different gauge restoring schemes.
In \tab{tab1} we give the cross sections for CC20 at Lep 2 and LC energies.
In \tab{tab2} we give the cross sections for CC20 at $E=800 \GeV$
with slightly modified selections. With all other cuts at their standard values,
in the second column the electron scattering angle is not 
allowed to be larger than
0.1 degree while in the third column the invariant mass of the $u\bar{d}$ pair
is required to be greater than $40 \GeV$\/. 

The IFL, FW, CM and  OA schemes agree
within $2\,\sigma$ in almost all cases. The IFLCC scheme agrees with all other
ones at Lep 2 energies but already at 800 $\GeV$ it overestimates the total
cross section by about 6\%. At $1.5 \TeV$ the error is almost a factor of two.
The results in  \tab{tab2} show that the discrepancy between the IFLCC scheme
and all the others decreases slightly to 5.6 \% if larger masses of the
$u\bar{d}$ pair are required. 
If instead smaller electron scattering angle are allowed the discrepancy
increases to about 9\%. This is a consequence of the fact that 
in the collinear region the neglected terms, 
 proportional to the fermion masses, are enhanced
by factors of order $\ord{m_f^2\gamma_{_W}(p^2)/(M_{_W}^2 m_e^2)}$
which can become very large at high energy even for typical light fermion
masses.

We conclude then that, even in the presence of non--conserved currents
i.e. of massive external fermions, the FW, CM and OA calculations give 
predictions
which are in agreement, within a few per mil, with the IFL scheme. 
This agreement with the
results of a fully self-consistent approach justifies from
a practical point of view the ongoing use of the FW, CM and OA schemes.
It should be remarked that for massless fermions 
it has been shown that at high energies, for the
total cross section of the process \processccten\ 
the full FL scheme deviates from the FW 
scheme and the IFL scheme by about 2\% at $1 \TeV$ increasing
to about 7\% at  $10 \TeV$ \cite{bhf2} mainly because of the running
of the couplings. As a consequence, it appears likely that calculations 
performed in the
IFL scheme with running couplings would be able to reproduce the complete
FL results with sufficient accuracy for most practical purposes.
Hitherto missing higher order QCD and bosonic contributions
could still conceivably produce significant corrections.

\section{Conclusions}

The Imaginary Part Fermion-Loop scheme, introduced in  Ref.~\cite{bhf1} 
for the gauge-invariant treatment of the finite-width effects of $W$ and $Z$
bosons, has been generalized so that it could be applied to processes with
massive external fermions. This involves  the  Dyson 
resummation of higher order imaginary contributions to the propagator which 
implies, in the unitary gauge, a modification of the $p^\mu p^\nu$ term in the 
numerator. From a numerical 
point of view we find no significant difference between
the IFL scheme and the FW, CM or OA schemes in the region most sensible to
U(1) gauge invariance.

\section*{Acknowledgments} 
This research has been partly supported by  
NSF Grant No. PHY-9722090 and by MURST.\\
We wish to thank G.~Passarino for several discussions on gauge invariance and related
issues. We also gratefully acknowledge the exchange of information and 
comparisons with other groups and in particular with
E.~Boos, M.~Dubinin, S.~Jadach and R.~Pittau.


\begin{thebibliography}{99}

\bibitem{berendschapovsky}W.~Beenakker, F.A.~Berends and A.P. Chapovsky,
hep-ph/9909472.

\bibitem{bhf1} 
E.N. Argyres et al., Phys.\ Lett.\ {\bf B358} (1995) 339.

\bibitem{baurzepp} U.~Baur and D.~Zeppenfeld,
Phys.\ Rev.\ Lett.\ {\bf 75} (1995) 1002.

\bibitem{bhf2}
W.~Beenakker  et al.,  Nucl.\ Phys. {\bf B500} (1997) 255.

\bibitem{Hoogland&vanoldenborgh}
J.~Hoogland and G.~J. van Oldenborgh, Phys.\ Lett. {\bf B402} (1997) 379.

\bibitem{GPnew} G.~Passarino, in preparation.

\bibitem{lmunu} Y.~Kurihara, D.~Perret-Gallix and Y.~Shimizu,
Phys.\ Lett. {\bf B349} (1995) 367.

\bibitem{revs} E. Accomando, CTP-TAMU-36/99, to appear in Proc. of the 
Int. Workshop on
        Linear Colliders, Sitges, 1999; 
A.~Ballestrero, DFTT 59/99, hep-ph/9911235;
 E.E.~Boos and M.N.~Dubinin, hep-ph/9909214.

\bibitem{iww} G.~Passarino, hep-ph/9810416.

\bibitem{wto} G.~Passarino, Comp. Phys. Comm. {\bf 97} (1996) 261.

\bibitem{com} A.~Pukhov et al.,  hep-ph/9908288.

\bibitem{grc} J.~Fujimoto et al., Comp. Phys. Comm. {\bf 100} (1997) 128.

\bibitem{kor} S.~Jadach et al., Comp. Phys. Comm. {\bf 119} (1999) 272. 

\bibitem{wph} E.~Accomando and  A.~Ballestrero,
Comp. Phys. Comm. {\bf 99} (1997) 270.


\bibitem{nextc} F.~Berends, A.~Kanaki, C.G.~Papadopoulos and R.~Pittau,
private communication.

\bibitem{swap} G.~Montagna, M.~Moretti, O.~Nicrosini, A.~Pallavicini and
F.~Piccinini, in preparation.

\bibitem{DDRW} A.~Denner, S.~Dittmaier, M.~Roth and D.~Wackeroth,
BI-TP 99/10, PSI--PR--99--12, hep-ph/9904472.

\bibitem{overall} U.~Baur, J. Vermaseren and D.~Zeppenfeld,
Nucl.\ Phys. {\bf B375} (1992) 3.


\end{thebibliography}
\end{document}